\documentstyle[12pt,fleqn]{article}
\def\beeq{\begin{equation}}
\def\eneq{\end{equation}}
\def\beeqa{\begin{eqnarray}}
\def\eneqa{\end{eqnarray}}

\addtocounter{section}{-1}
\begin{document}

\begin{center}

\vspace{2cm}

{\large {\bf {Electronic states of metallic and semiconducting\\
carbon nanotubes with bond and site disorder
} } }

\vspace{1cm}

{\rm Kikuo Harigaya\footnote[1]{E-mail address: 
\verb+harigaya@etl.go.jp+; URL: 
\verb+http://www.etl.go.jp/+\~{}\verb+harigaya/+}}

\vspace{1cm}

{\sl Physical Science Division,
Electrotechnical Laboratory,\\ 
Umezono 1-1-4, Tsukuba 305-8568, 
Japan}\footnote[2]{Corresponding address}\\
{\sl National Institute of Materials and Chemical Research,\\ 
Higashi 1-1, Tsukuba 305-8565, Japan}\\
{\sl Kanazawa Institute of Technology,\\
Ohgigaoka 7-1, Nonoichi 921-8501, Japan}

\vspace{1cm}
(Received~~~~~~~~~~~~~~~~~~~~~~~~~~~~~~~~~~~)
\end{center}

\vspace{1cm}

\noindent
{\bf Abstract}\\
Disorder effects on the density of states in carbon nanotubes 
are analyzed by a tight binding model with Gaussian bond or 
site disorder.  Metallic armchair and semiconducting zigzag 
nanotubes are investigated.  In the strong disorder limit, 
the conduction and valence band states merge, and a finite 
density of states appears at the Fermi energy in both of 
metallic and semiconducting carbon nanotubes.  The bond 
disorder gives rise to a huge density of states at the Fermi 
energy differently from that of the site disorder case.
Consequences for experiments are discussed.

\vspace{1cm}

\pagebreak

\section{Introduction}

Recently, carbon nanotubes with cylindrical graphite 
structures have been intensively investigated.  Many interesting 
experimental as well as theoretical researches have been 
performed (see reviews [1,2] for example), and the fundamental
metallic and semiconducting behaviors of single wall nanotubes
predicted by theories [3-8] have been clarified in tunneling 
spectroscopy experiments [9,10].

In the preceding work [11], we have studied changes of the 
density of states and electronic conduction in metallic 
carbon nanotubes with using a tight binding model with 
Gaussian bond disorder.  Metallic armchair and zigzag 
nanotubes have been investigated.  We have obtained a 
conductance which becomes smaller by the factor $1/2 - 1/3$ 
from that of the clean nanotube.  We have also found that 
suppression of electronic conductance around the Fermi 
energy due to disorder is smaller than that of the inner 
valence (and conduction) band states, as a consequence of 
the extended nature of electronic states around the Fermi 
energy between the valence and conduction bands, and is a 
property typical of the electronic structures of metallic 
carbon nanotubes.

The purpose of this paper is to give calculations on 
semiconducting carbon nanotubes newly, which have not 
been studied in the previous report [11].  Site as well 
as bond disorder effects on the density of states of 
metallic and semiconducting nanotubes are considered.  
The bond and site disorder model has been used in the
discussion of polarons in doped C$_{60}$ [12], too.
We will discuss the following.  In the strong disorder 
limit, the conduction and valence band states merge, 
and a finite density of states appears at the Fermi 
energy in both of metallic and semiconducting carbon 
nanotubes.  A huge density of states appears at the 
Fermi energy in the bond disorder case.  This property 
does not occur for the site disorder case.  Consequences
for experiments are discussed.

\section{Model}

We will study the following model for carbon nanotubes 
with bond disorder:
\beeq
H = - t \sum_{\langle i,j \rangle, \sigma} 
(c_{i,\sigma}^\dagger c_{j,\sigma} + {\rm h.c.})
+ \sum_{\langle i,j \rangle, \sigma} \delta t_{i,j}
(c_{i,\sigma}^\dagger c_{j,\sigma} + {\rm h.c.}).
\eneq
The first term is the tight binding model with the nearest 
neighbor hopping interaction $t$; the sum is taken over 
neighboring pairs of lattice sites $\langle i,j \rangle$ 
and spin $\sigma$; $c_{j,\sigma}$ is an annihilation operator
of an electron with spin $\sigma$ at the site $i$.  The 
second term is the bond disorder model, and the hopping 
interaction $\delta t_{i,j}$ obeys the Gaussian distribution 
function
\beeq
P(\delta t)=\frac{1}{\sqrt{4\pi}t_s}
{\rm exp}[-\frac{1}{2}(\frac{\delta t}{t_s})^2]
\eneq
with the strength $t_s$.

In the site disorder case, the model hamiltonian is:
\beeq
H = - t \sum_{\langle i,j \rangle, \sigma} 
(c_{i,\sigma}^\dagger c_{j,\sigma} + {\rm h.c.})
+ \sum_{i, \sigma} \delta U_i c_{i,\sigma}^\dagger c_{i,\sigma},
\eneq
where $\delta U_i$ is the onsite potential due to site 
disorder whose distribution is the Gaussian function with 
the strength $U_s$.

A finite system with the quite large system size $N$ of 
metallic carbon nanotubes is diagonalized numerically.  
In this paper, we take $N=4000$ for the (5,5) metallic
nanotube, and for the (10,0) semiconducting zigzag tube.
The quantity $t_s$ is changed within $0 \leq t_s \leq 1t$,
and $U_s$ is varied within $0 \leq U_s \leq 1t$.  All the
quantities with the dimension of energy are measured in 
units of the hopping integral $t$ ($\sim 2$ eV) in this 
paper.

\section{Bond disorder effects}

Figure 1 shows the density of states (DOS) of the metallic
(5,5) nanotube with bond disorder.  Figures 1 (a), (b), and (c) 
show the weak disorder case with $t_s = 0.1t$, the middle
strength case $t_s=0.5t$, and the strong limit case
$t_s = 1.0t$, respectively.  When $t_s = 0.1t$, the DOS
retains its one-dimensional singularities.  The magnitude
of the constant DOS around the Fermi energy $E=0$ is near to 
that of the clean system.  As $t_s$ increases, the DOS of 
the valence and conduction band states broadens.  In the case 
of $t_s = 0.5t$ (Fig. 1 (b)), the DOS is very broad and loses 
the one-dimensional characters.  The magnitude of the DOS at the
Fermi energy does not decrease, but it enhances from
that of the clean system.  This is due to the large 
overlap of the original valence and conduction band
states.  In the strong $t_s$ case (Fig. 1 (c)), we find
{\sl a surprising sharp peak} at the Fermi energy.  This is
not an accidental result of the calculation.  The peak
value of the DOS is a few times larger than that of
the bodies of the valence and conduction band states
at $t_s = 1.0t$.  The DOS at the Fermi energy is more than 
ten times larger at $t_s = 1.0t$ than that of the clean 
system $t_s = 0$.

Next, we consider the DOS of the semiconducting (10,0)
nanotube with bond disorder.  Fig. 2 shows the calculated
DOS for the three disorder strengths.  In the weak disorder
case $t_s = 0.1t$ (Fig. 2 (a)), the DOS well retains
its one-dimensional features.  However, as the disorder
strength becomes larger, the energy gap decreases.
And, a finite density of states is present at the
center of energy as shown for the case $t_s = 0.5t$ 
(Fig. 2 (b)).  This merging of the valence and conduction 
band states might mean a kind of insulator-semimetal 
transitions.  Here, the transition to a semimetal simply 
refers to an appearance of a finite density of states 
around the Fermi energy in the presence of disorder potentials.
In the strong disorder case (Fig. 2 (c)), we again find
{\sl a huge density of states} at the origin of energy.  Such
the common result between the metallic and semiconducting
nanotubes might mean that the feature at the strong disorder
limit does not depend on whether the clean system is 
metallic or not.

It is interesting to look at the disorder strength 
dependence of the DOS, and compare the two cases of
the metallic and semiconducting carbon nanotubes
(Figs. 1 and 2).  Figure 3 show the variations as
a function of the bond disorder strength $t_s$. 
Squares are for the metallic (5,5) nanotube, and
crosses are for the semiconducting (10,0) nanotube.
In the weak disorder region ($t_s < 0.4 t$)
of the (5,5) nanotube, the density of states is 
nearly constant due to the flatness around the Fermi 
energy (Fig. 1 (a)).  After merging the valence and
conduction band states ($t_s > 0.5t$), the DOS 
becomes more than one order of magnitudes larger
owing to the appearance of the huge peak (Fig. 1 (c)).  
The DOS even becomes about ten times larger than 
that of the clean metallic system.  For the semiconducting 
(10,0) nanotube, the DOS at the gap center is negligible
when $t_s < 0.4t$.   The development of the density of states 
after the insulator-semimetal transition is very similar
to that of the metallic (5,5) system.  The similarity
is quantitatively as well as qualitatively.  Such
the similarity indicates that the result in the strong
$t_s$ limit is general as the bond disorder effects,
not depending on whether the clean system is metallic
or not.

\section{Site disorder effects}

In this section, we turn to the site disorder effects.
As in the previous section, results of the (5,5) metallic 
nanotube are reported in Fig. 4, and those of the (10,0)
semiconducting nanotube are reported in Fig. 5.  The 
disorder strengths are $U_s = 0.1t$ (Figs. 4 (a) and 5 (a)),
$U_s = 0.5t$ (Figs. 4 (b) and 5 (b)), and $U_s = 1.0t$
(Figs. 4 (c) and 5 (c)), respectively.  In the weak disorder
case ($U_s = 0.1t$), one-dimensional features in the entire
density of states can be seen clearly.  As the strength
becomes larger, the merging of the valence and conduction
bands occurs (Figs. 4 (b) and 5 (b)).  A kind of insulator-semimetal 
transition is found for Fig. 5, as in the bond disorder case.
Then, the entire DOS becomes very broad in the strong 
disorder limit (Figs. 4 (c) and 5 (c)).  We thus find 
{\sl a qualitative difference} from that of the bond 
disorder case: the absence of the huge peak around the 
energy center.  Rather, the site impurity effects on the 
DOS seem usual as general theories of disorder predict.

Finally, Figure 6 shows the disorder strength $U_s$ 
dependence of the DOS at the Fermi energy.  Squares
are for the (5,5) nanotube, and the crosses are for
the (10,0) nanotube.  The nearly constant behavior
in the weak disorder region of the (5,5) nanotube case 
is similar to that of the bond disorder effect.  In the 
strong disorder ($U_s \sim 1.0t$), the DOS enhances its 
magnitude more or less from that of the weak disorder.
However, the order of the magnitudes is the same.
In the (10,0) tube case, the transition to the
semimetal occurs at $U_s = 0.6t$.  The DOS of the strong
$U_s$ region is like that of the squares.  The magnitude 
of the DOS is of the same order, even if we do calculations for
the disorder strengths upto $U_s \sim 5t$.  Therefore,
we conclude that a huge density of states does
not appear in the site disorder case differently
from the bond disorder effects.

The bond and site disorder effects in the weak strength
region are similar mutually as we expect from general 
theory of disorder systems.  However, there is a
remarkable difference in the strong disorder limit.
This might come from the difference of how the disorder
potential works.  In the literature [13], a one dimensional
system with bond disorder has been analyzed.  The analytic
formula shows the logarithmic divergence of the DOS at
the Fermi energy.  The similar effect might take place
in the present numerical calculations of carbon nanotubes
with bond disorder.

In the real carbon nanotubes, bond as well as site
disorder can be present simultaneously, even though
the two strengths might be different.  In cases
with stronger bond disorder and weak site disorder, 
we might be able to find a large density of states 
at the energy center.  Here, an originally semiconducting 
system could exhibit a finite density of states. 
Such an observation seems interesting if some kinds of 
measurements are possible.

\section{Summary}

In summary, we have investigated disorder effects on the 
density of states in metallic and semiconducting carbon 
nanotubes, using a tight binding model with Gaussian bond 
or site disorder.  Armchair and zigzag nanotubes have 
been considered.  The systems with weak disorder are
similar in both disorder cases.  However, as the disorder
becomes stronger the conduction and valence band states 
merge, and a finite density of states appears at the Fermi 
energy in both of metallic and semiconducting carbon 
nanotubes.  The bond disorder gives rise to a huge density 
of states at the Fermi energy differently from that 
of the site disorder case.

\mbox{}

\begin{flushleft}
{\bf Acknowledgements}
\end{flushleft}

\noindent
Useful discussion with the members of Condensed Matter
Theory Group\\
(\verb+http://www.etl.go.jp/+\~{}\verb+theory/+),
Electrotechnical Laboratory is acknowledged.  Numerical 
calculations have been performed on the DEC AlphaServer 
of Research Information Processing System Center (RIPS), 
Agency of Industrial Science and Technology (AIST), Japan.

\pagebreak
\begin{flushleft}
{\bf References}
\end{flushleft}

\noindent
$[1]$ M. S. Dresselhaus, G. Dresselhaus, and P. C. Eklund,
``Science of Fullerenes and Carbon Nanotubes",
(Academic Press, San Diego, 1996).\\
$[2]$ R. Saito, G. Dresselhaus, and M. S. Dresselhaus,
``Physical Properties of Carbon Nanotubes",
(Imperial College Press. London, 1998).\\
$[3]$ J. W. Mintmire, B. I. Dunlap, and C. T. White,
Phys. Rev. Lett. {\bf 68}, 631 (1992).\\
$[4]$ N. Hamada, S. Sawada, and A. Oshiyama,
Phys. Rev. Lett. {\bf 68}, 1579 (1992).\\
$[5]$ R. Saito, M. Fujita, G. Dresselhaus, and M. S. Dresselhaus,
Appl. Phys. Lett. {\bf 60}, 2204 (1992).\\
$[6]$ K. Tanaka, K. Okahara, M. Okada, and T. Yamabe,
Chem. Phys. Lett. {\bf 193}, 101 (1992).\\
$[7]$ K. Harigaya, Phys. Rev. B {\bf 45}, 12071 (1992).\\
$[8]$ K. Harigaya and M. Fujita, Phys. Rev. B {\bf 47},
16563 (1993).\\
$[9]$ J. W. G. Wild\"{o}er, L. C. Venema, A. G. Rinzler,
R. E. Smalley, and C Dekker. Nature {\bf 391}, 59 (1998).\\
$[10]$ T. W. Odom, J. L. Huang, P. Kim, and C. M. Lieber,
Nature {\bf 391}, 62 (1998).\\
$[11]$ K. Harigaya, cond-mat/9810341.\\
$[12]$ K. Harigaya, Phys. Rev. B {\bf 48}, 2765 (1993).\\
$[13]$ T. P. Eggarter and R. Riedinger, Phys. Rev. B
{\bf 18}, 569 (1978).\\

\pagebreak

\begin{flushleft}
{\bf Figure Captions}
\end{flushleft}

\mbox{}

\noindent
Fig. 1. Density of states (DOS) of the (5,5) nanotube with
bond disorder.  Figs. (a), (b), and (c) show the cases
with $t_s = 0.1t$, $0.5t$, and $1.0t$, respectively.

\mbox{}

\noindent
Fig. 2. Density of states (DOS) of the (10,0) nanotube with
bond disorder.  Figs. (a), (b), and (c) show the cases
with $t_s = 0.1t$, $0.5t$, and $1.0t$, respectively.

\mbox{}

\noindent
Fig. 3. Density of states (DOS) at the Fermi energy $E=0$
as a function of $t_s$.  Squares and crosses show the results
of the (5,5) and (10,0) nanotubes, respectively.

\mbox{}

\noindent
Fig. 4. Density of states (DOS) of the (5,5) nanotube with
site disorder.  Figs. (a), (b), and (c) show the cases
with $U_s = 0.1t$, $0.5t$, and $1.0t$, respectively.

\mbox{}

\noindent
Fig. 5. Density of states (DOS) of the (10,0) nanotube with
site disorder.  Figs. (a), (b), and (c) show the cases
with $U_s = 0.1t$, $0.5t$, and $1.0t$, respectively.

\mbox{}

\noindent
Fig. 6. Density of states (DOS) at the Fermi energy $E=0$
as a function of $U_s$.  Squares and crosses show the results
of the (5,5) and (10,0) nanotubes, respectively.

\end{document}